\documentclass{acm_proc_article-sp}

\usepackage[noend]{algorithmic}
\usepackage{algorithm}
\usepackage{float}
\newdef{definition}{Definition}
\newcommand{\byd}{\stackrel{\mbox{\tiny def}}{=}}
\def\calL{\mathcal{L}}

\begin{document}
\title{Applying FCA toolbox to Software Testing}
\subtitle{}
\numberofauthors{1}
\author{
\alignauthor
Fedor Strok\\
  \affaddr{Yandex}\\
  \affaddr{16 Leo Tolstoy St. Moscow, Russia}\\
  \email{fdrstrok@yandex-team.ru}\\
  \affaddr{Faculty for Computer Science}\\
  \affaddr{National Research University Higher School of Economics}\\
  \affaddr{Kochnovskiy Proezd, 3, Moscow, Russia}\\
  \email{fdr.strok@gmail.com}
}
\maketitle
\begin{abstract}
Software testing uses wide range of different tools to enhance the complicated process of defining quality of the system under test. Formal Concept Analysis (FCA) provides us with algorithms of deriving formal ontology from a set of objects and their attributes.  With the use of FCA we can considerably improve the efficiency of test case derivation. Moreover, an FCA-based machine learning system supports the analysis of regression testing results.
\end{abstract}

\section{Introduction}
Software testing aims at understanding the risks of software implementation and providing proper quality of the system under test. One of fundamental problems is that testing under all possible combinations of inputs is not feasible. This holds for functional testing, while non-functional aspects (performance, compatibility, etc.) are left apart. For business critical applications black-box testing is a widely-used approach. It examines the functionality of a system under test without dealing with its implementation. It is important to notice that static testing involves verification, while dynamic testing involves validation.

One of the biggest challenges in black-box dynamic testing is choosing the proper approach to enumerate all possible cases. Domain testing helps quality assurance engineer to define classes of input values, that are crucial to test. The naive way is just to check for possible combinations of such parameters, and it immediately leads to exponential complexity. Even in the case of 6 boolean parameters an expert has to check 64 combinations.

A possible solution is to generate test-cases in manual way. A software test engineer considers all possible cases ordered in a certain way. The main risk is to lose some information behind the cases. Actually, it could be rather exhausting to cover all possibilities for a large number of parameters and not to skip some scenarios.

A popular alternative way is pairwise testing~\cite{cz}, or its generalization, n-wise testing. We define parameters and domains, and pass them like a model into a black-box algorithm~\cite{bach}, which gives us a set of test-cases satisfying a certain condition (for each pair of input parameters all possible discrete combinations of those parameters are tested). In general case it can produce different cases in different runs, while it can be fixed by passing a random seed to it. The main advantage of the approach is its insensitivity to parameter values. However, it can be a rather computationally demanding task.

For quality assurance engineers it is sufficient to have knowledge about possible behavior of the program. Usually, there are dependencies between input parameters. A natural form to express such dependencies in mathematical terms is implication, a statement of the form: `if ..., then ...'. The `if'-part is called premise, and the `then' is called conclusion. Consideration of parameter's interdependence can decrease the complexity of result test-cases in terms of their quantity, by excluding some of possible parameter combinations.

Proposed approach to test-case generation is focused on implications. We use an approach based on Formal Concept Analysis (FCA)\cite{gw99}. FCA provides software engineers with a tool for exploring the domain of interest in semi-automatic way. The algorithm outputs sound and complete description of the problem, but it is still highly dependent on the accuracy of expert's answers. Surveys on FCA techniques and applications can be found in~\cite{pk}.   FCA-based approaches have already been applied in software engineering, e.g., for inference of class hierarchies ~\cite{sn}, class design ~\cite{gv}, refactoring ~\cite{mh}. Another application of FCA technique is analysis of object-oriented approach for a given system ~\cite{tc}. The questions of mapping lines of source code to the functionality from requirements is of crucial importance for big systems ~\cite{pm}. 

The rest of the paper is organized as follows: in Section 2 we introduce basic notions of Formal Concept Analysis. In Section 3 we focus on the procedure of attribute exploration. Section 4 provides example of attribute exploration in the field of positive integers. We make conclusions in Section 5.

\section{Formal Concept Analysis}
Formal Concept Analysis~\cite{gw99} (FCA) provides  mathematical technique for deriving applied ontologies from data. FCA relies on lattice and order theories~\cite{dp02} Numerous applications are found in the field of machine learning, data mining, text mining and biology, see~\cite{pk,pk2}.
\subsection{Main Definitions}
 \begin{definition}
A formal context $K$ is  a triple $K:=(G,M,I)$, where G denotes a set of objects, M is a set of attributes, and $I\subseteq G\times M$ is a binary relation between $G$ and $M$.
 \end{definition}
It can be interpreted in the following way: for objects in $G$ there exists a description in terms of attributes in $M$, and relation $I$ reflects that an object has an attribute:  $(g,m)\in I\iff $ object $g$ possesses~$m$.

An example of a formal context is provided below:
\begin{center}
\begin{tabular}{|c|cccc|}
  \hline
   G $\setminus$ M & a & b & c & d\\
  \hline
   \begin{picture}(10,10)
      \put(0,0){\line(1,0){10}}
      \put(0,0){\line(1,2){5}}
      \put(10,0){\line(-1,2){5}}
      \end{picture} &$\times$ &  & & $\times$\\
  \begin{picture}(10,10)
      \put(0,0){\line(1,0){10}}
      \put(0,0){\line(0,1){10}}
      \put(0,10){\line(1,-1){10}}
      \end{picture} &$\times$ &  & $\times$ &\\
   \begin{picture}(20,10)
      \put(0,0){\line(1,0){20}}
      \put(0,0){\line(0,1){10}}
      \put(0,10){\line(1,0){20}}
      \put(20,0){\line(0,1){10}}
      \end{picture} &  &$\times$ & $\times$ &\\
   \begin{picture}(10,10)
      \put(0,0){\line(1,0){10}}
      \put(0,0){\line(0,1){10}}
      \put(0,10){\line(1,0){10}}
      \put(10,0){\line(0,1){10}}
      \end{picture} & & $\times$ &$\times$ & $\times$\\

  \hline
  \end{tabular}
\end{center}
\vspace*{.6cm}

  \qquad Objects:\hspace*{3cm} Attributes: \par
  \vspace*{-.4cm}
  \begin{tabular}{p{4cm}p{4cm}}
  \begin{itemize}
  \item[\textbf{1}] -- equilateral triangle
  \item[\textbf{2}] -- right triange,
  \item[\textbf{3}] -- rectangle,
  \item[\textbf{4}] -- square,
  \end{itemize} &
  \begin{itemize}
  \item[\textbf{a}] -- 3 vertices,
  \item[\textbf{b}] -- 4 vertices,
  \item[\textbf{c}] -- has a right angle,
  \item[\textbf{d}] -- all sides are equal
  \end{itemize}
  \end{tabular}
\vspace*{.6cm}

For a given context two following mappings are considered:
\begin{itemize}
\item[] $ \varphi\colon 2^G\to 2^M \, \varphi(A) \byd \{m\in M\mid gIm \ \mbox{for all\ } g\in A\}. $
\item[] $ \psi\colon 2^M\to 2^G \, \quad \psi(B) \byd \{g\in G\mid gIm \ \mbox{for all\ } m\in B\}. $
\end{itemize}

For all $A_1, A_2\subseteq G$, $B_1, B_2\subseteq M$
\begin{enumerate}
\item $A_1\subseteq A_2 \Rightarrow \varphi(A_2) \subseteq \varphi(A_1)$
\item $B_1\subseteq B_2 \Rightarrow \psi(B_2) \subseteq \psi(B_1)$
\item $A_1\subseteq \psi\varphi (A_1) \mbox{ and} B_1\subseteq \varphi\psi(B_1)$
\end{enumerate}

\begin{definition}
Mappings $\varphi$ and $\psi$, satisfying properties 1-3 above, define a Galois connection between $(2^G,\subseteq)$ and $(2^M,\subseteq)$, which means:
$\varphi(A)\subseteq B  \Leftrightarrow \psi(B)\subseteq ~A$
\end{definition}
Traditionally, notation  $(\cdot)^{\prime}$ is used instead of $\varphi$ and $\psi$. $(\cdot)^{\prime\prime}$ stands
both for $\varphi\circ\psi$ and $\psi\circ\varphi$ (depending on the argument).
For arbitrary $A\subseteq G$, $B\subseteq M$
$$
A^{\prime} \byd \{m\in M\mid gIm \ \mbox{for all\ } g\in A\},\quad $$
$$B^{\prime}
\byd \{g\in G\mid gIm \ \mbox{for all\ } m\in B\}.
$$
\begin{definition}
\emph{(Formal) concept} is a pair $(A,B)$:
${A\subseteq G}$, ${B\subseteq M}$, ${A^{\prime} = B}$, $B^{\prime} = ~A.$
\end{definition}
In the example with geometric  figures a pair $(\{3,4\},\{b,c\})$ is a formal concept.
For a formal context $(G,M,I)$, $A, A_1, A_2\subseteq G$ - set of objects, $B\subseteq M$ - set of attributes, the following statements hold for operation ($\cdot$)':
\begin{enumerate}
\item $A_1\subseteq A_2 \Rightarrow  A'_2\subseteq A'_1$
\item $A_1 \subseteq A_2 \Rightarrow  A^{''}_1 \subseteq A^{''}_2$
\item $A\subseteq A''$
\item $A''' = A'$ and $A'''' = A''$
\item $(A_1 \cup A_2)' = A'_1 \cap A'_2$
\item $A\subseteq  B'\Leftrightarrow B\subseteq A' \Leftrightarrow A\times B \subseteq I$
\end{enumerate}

\begin{definition}
\emph{Closure operator} on set $G$ is a mapping $\gamma \colon {\cal P}(G)\to {\cal P}(G)$,
which maps every $X\subseteq G$ to \emph{closure} $\gamma X\subseteq G$, under the following conditions:
\begin{enumerate}
\item ${\gamma{\gamma X}} = {\gamma X}$ (\emph{idempotence})
\item $X\subseteq {\gamma X}$ (\emph{extensivity})
\item $X\subseteq Y\Rightarrow {\gamma X}\subseteq {\gamma Y}$ (\emph{monotonicity})
\end{enumerate}
\end{definition}

\begin{definition}
\emph{Implication} $A\to B$, where $A, B\subseteq M$, takes place if
$A'\subseteq B'$, in other words if each object having
$A$ also has all attributes from $B$.
\end{definition}
Implications comply with Armstrong axioms:
$${{}\over{X\to X}}\hfill (1)$$
        $${{X\to Y}\over{X\cup Z\to Y}}\hfill (2)$$
        $${{X\to Y, Y\cup Z\to W}\over{X\cup Z\to W}}\hfill (3)$$

\section{Lattice diagrams}
One of advantages of building lattices of formal concepts is to get effective navigation from more general concepts to more specific. For example, the line diagram for context with figures from previous section is shown in Fig. 1.
\begin{figure}
\label{figures_lat}
\includegraphics[width=8cm,height=7cm]{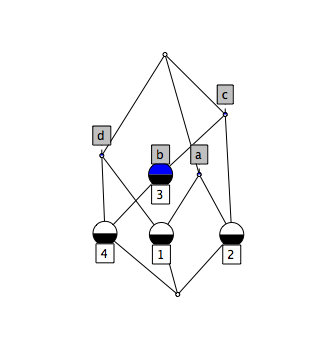}
\caption{Line diagram for context with geometric figures}
\end{figure}
That property could be beneficial in two ways: regression testing, system description.

First of all, for the context of regression testing it implements the algorithm to determine the classes of tests that fail. It could be considered in the following way. Let us define set of attributes for all tests that are being run.
\begin{center}
\begin{tabular}{|c|cccc|}
  \hline
   G $\setminus$ M & failed & https & login & messages\\
  \hline
   1 &$\times$ & $\times$ & $\times$ & \\
  2 & & $\times$  & &$\times$\\
   3 &  $\times$ & & $\times$ &\\
   4 & &  & & $\times$\\

  \hline
  \end{tabular}
\end{center}
\vspace*{.6cm}

\begin{figure}
\includegraphics[width=8cm]{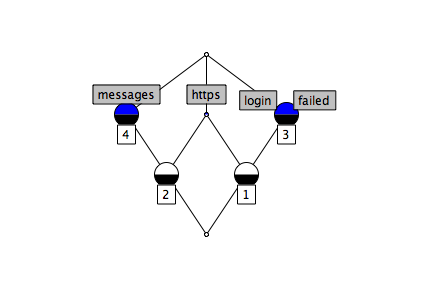}
\caption{Test run: Line diagram of all tests}
\end{figure}
\begin{figure}
\includegraphics[width=8cm]{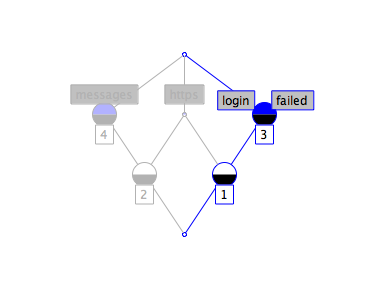}
\caption{Test run: line diagram of Concepts with failed=True highlighted}
\end{figure}
Then we can view the line diagram of all the tests during the run. See Fig. 2. The most interesting formal concepts are those with attribute ``failed=True'', since the simplified result could be treated as either success or fail. Now we can notice that ``messages'' part is ok, while login module breaks totally, and there is one problem with ``https''. The convenient strategy for making such meta-reports is to filter the set of all tests which have ``failed=True'' flag. Then we can build concept lattice for obtained filtered formal context. Moreover, we could build just the top part of concept lattice, since it will have most general attributes.

The second important application of formal concepts in big systems is analysis of dependent features. Since modern software development process usually is organized in agile manner, it is important to view the effect of developed features on the overall system, and examine connected components. It could be fruitful to use test cases from features with common functionality. To track such features and to have user-friendly navigation we can use FCA-based techniques. We need to have the list of features and set of tags as general components description. A small example is provided below. We can build formal concept lattice for the context describing the features. Then we can search obtained graph, finding not only features with the same set of tags, but also more general and more specific ones. The initial lattice is depicted in Fig. 4. If we want to analyze the feature with tags ``https'' and ``login'' we see that it forms formal concept (\{f1,f3,f5\},\{\})
\begin{center}
\begin{tabular}{|c|ccccc|}
  \hline
   G $\setminus$ M & billing & https & login & messages& static\\
  \hline
   f1 &$\times$ & $\times$ & $\times$ & &\\
  f2 & & $\times$  & &$\times$&\\
   f3 &  $\times$ & $\times$& $\times$ &&$\times$\\
   f4 & &  &$\times$ & $\times$&\\
   f5 & &$\times$  & $\times$& &$\times$\\
  \hline
  \end{tabular}
\end{center}
\begin{figure}
\includegraphics[width=8cm]{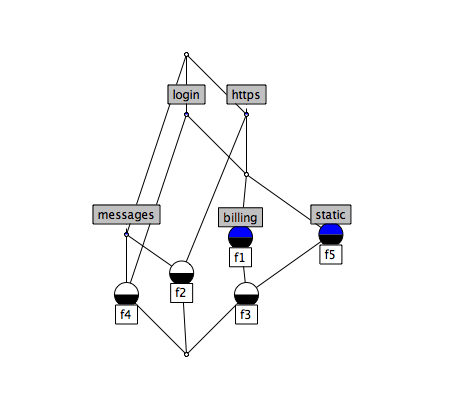}
\caption{Features description}
\end{figure}
\begin{figure}
\includegraphics[width=8cm]{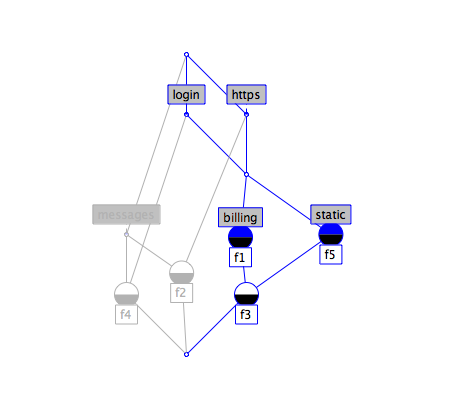}
\caption{Features description: concept with ``https'' and ``login''}
\end{figure}

\section{Feature Impact Analysis}
One of the most important steps of testing software is to determine the functionality that would be changed. Usually, all activities concerning the development process are stored in issue trackers. It is important to reuse the experience of previous features while developing new functionality. We can think of two dimensions of analysis of related features. Firstly, issue trackers typically provide the possibility to link tickets and in this way we can draw graph of similar features. Secondly, we can benefit from tags that are used to mark system parts. So, example from the previous section could be studied in alternative way: we can browse features to find the closest one in terms of functional units. We can find out which cases were obligatory in this case and simplify test-case design process.

\section{Attribute Exploration}

Attribute exploration is a well-known FCA-based discovery technique. The main idea is to explore the implications in the object domain in a semi-automatic manner. According to Attribute Exploration a domain expert answers specific questions about possible implicational dependencies in his domain. Questions are provided in the form of attribute implications, asking whether they are true or false. If the answer is true, the implication is added to the knowledge base. In the case of answer ``false'', the expert is asked to provide a counterexample violating the proposed dependency.

In other words, the exploration algorithm explores all possible combinations of a given attribute set. It is typical that objects in this field of knowledge are too difficult to enumerate them. So the algorithm starts with a set of examples, then it computes canonical base of implications for the provided formal context. Then the domain expert is asked if the computed implications are valid. If it is true, than the existing context represents all possible combinations in the domain. Otherwise, the expert has to provide a counterexample to implications, which should be added to the context as new objects, and then the implication base is recomputed.

The general strategy is quite intuitive: we start exploring the domain with some knowledge of typical examples and implications. To extend the knowledge base we either add an implication, or provide another example that violates currently studied implication. The main advantage of this approach is that generation of dependencies (implications) is performed algorithmically.
\algsetup{indent=2em}
\renewcommand{\algorithmicrequire}{\textbf{Input:}}
\renewcommand{\algorithmicensure}{\textbf{Output:}}
\renewcommand{\algorithmicprint}{\textbf{output}}
\begin{algorithm}[H]
    \caption{\textsc{Next Closure}($A$, $M$, $\calL$)}
    \label{algo:nextclosure}
    \begin{algorithmic}
        \REQUIRE Closure operator $X\mapsto \calL(X)$ on attribute set $M$ and subset $A \subseteq M$.
        \ENSURE lectically next closed itemset $A$.
        \FORALL{$m \in M$ in reverse order}
            \IF{$m \in A$}
                \STATE $A := A \setminus \{m\}$
            \ELSE
                \STATE $B := \calL(A \cup \{m\})$
                \IF{$B \setminus A$ does not contain elemens $< m$}
                    \RETURN $B$
                \ENDIF
            \ENDIF
        \ENDFOR
        \RETURN $\bot$
    \end{algorithmic}
\end{algorithm}
\begin{algorithm}[H]
    \caption{\textsc{Attribute Exploration}}
    \label{algo:atribute_exploration}
    \begin{algorithmic}
        \REQUIRE A subcontext $(E,M,J=I \cap E\times M)$ of $(G,M,I)$, possibly empty.
        \REQUIRE Interactive: confirm that $A=B^{\prime\prime}$ in a formal context $(G,M,I)$, $M$ finite, or give an object showing that $A\neq B^{\prime\prime}$.
        \ENSURE The canonical base $\calL$ of $(G,M,I)$ and a possibly enlarged subcontext $(E,M,J=I\cap E\times M)$ with the same canonical base.
        \STATE $\calL := \emptyset$
        \STATE $A := \emptyset$
        \WHILE{$A \neq M $}
            \WHILE{$A\neq A^{JJ}$}
                \IF{$A^{JJ}=A^{II}$}
                  \STATE $\calL := \calL\cup \{A\to A^{JJ}\}$
                  \STATE exit while
                \ELSE
                  \STATE extend $E$ by some object $g\in A^I\setminus A^{JJI}$
                \ENDIF
            \ENDWHILE
            \STATE $A:= NextClosure(A,M,\calL)$
        \ENDWHILE
        \RETURN $\calL, (E,M,J)$
    \end{algorithmic}
\end{algorithm}

\section{Practical examples}
\subsection{Numbers}
Let us consider the domain of natural numbers~\cite{ae}. As a set of possible attributes we can consider the following ones: even (2*n), odd (2*n+1), divisible\_by\_three (3*n), prime (has no positive divisors other than 1 and itself), factorial (is a factorial of a positive number). We start from the empty set of objects. The canonical base for such context is $\emptyset \to M$.
So we get a question:

$=> even, factorial, divided\_by\_three, odd, prime$\\
Is the following implication valid?\\

Obviously, not all numbers have all attributes. At least we can consider number 2, which is $even, factorial, prime$. We add 2 to our context and the base is recomputed.
\begin{center}
\begin{tabular}{|c|c|c|c|c|c|}
  \hline
  G $\setminus$ M & even & factorial & divided\_by\_three & odd & prime\\
\hline
  2 & $\times$ & $\times$ &  & & $\times$ \\
  \hline
\end{tabular}
\end{center}

$ => even, factorial, prime$\\
Is the following implication valid?\\
Now we can consider number 5, which is $prime, odd$
\begin{center}
\begin{tabular}{|c|c|c|c|c|c|}
  \hline
  G $\setminus$ M & even & factorial & divided\_by\_three & odd & prime\\
\hline
  2 & $\times$ & $\times$ &  & & $\times$ \\
  5 &  & &  & $\times$ & $\times$ \\
  \hline
\end{tabular}
\end{center}
$ => prime$\\
Is the following implication valid?\\
Now we are about to either say that all numbers are prime, or provide a non-prime number, e.g. 6
\begin{center}
\begin{tabular}{|c|c|c|c|c|c|}
  \hline
  G $\setminus$ M & even & factorial & divided\_by\_three & odd & prime\\
\hline
  2 & $\times$ & $\times$ &  & & $\times$ \\
  5 &  & &  & $\times$ & $\times$ \\
  6 & $\times$ & $\times$ & $\times$ & & \\
  \hline
\end{tabular}
\end{center}
$factorial => even$\\

Is the following implication valid?\\
Now we have both 2 and 6, which are simultaneously even and factorial. There is a counterexample, we should find a number that is factorial, but not even, which is 1.
\begin{center}
\begin{tabular}{|c|c|c|c|c|c|}
  \hline
  G $\setminus$ M & even & factorial & divided\_by\_three & odd & prime\\
\hline
  2 & $\times$ & $\times$ &  & & $\times$ \\
  5 &  & &  & $\times$ & $\times$ \\
  6 & $\times$ & $\times$ & $\times$ & & \\
  1 & & $\times$ & & $\times$ & $\times$ \\
  \hline
\end{tabular}
\end{center}
$odd => prime$\\
Is the following implication valid?\\
No, it does not hold for number 9.
\begin{center}
\begin{tabular}{|c|c|c|c|c|c|}
  \hline
  G $\setminus$ M & even & factorial & divided\_by\_three & odd & prime\\
\hline
  2 & $\times$ & $\times$ &  & & $\times$ \\
  5 &  & &  & $\times$ & $\times$ \\
  6 & $\times$ & $\times$ & $\times$ & & \\
  1 & & $\times$ & & $\times$ & $\times$ \\
  9 & & & $\times$ & $\times$ & \\
  \hline
\end{tabular}
\end{center}
$factorial, odd => prime$\\
Is the following implication valid?\\
We have the only number 1 which is factorial and odd, and it is also prime.\\
$factorial, divided\_by\_three => even$\\
Is the following implication valid?\\
Now we have to recall what is implication? The only case where implication does not hold is when premise is true and conclusion is false. The least factorial which is divided\_by\_three is 6, which is already even.\\
$prime, divided\_by\_three => even, factorial, odd$\\
Is the following implication valid?\\
We have number 3, which is just odd.
\begin{center}
\begin{tabular}{|c|c|c|c|c|c|}
  \hline
  G $\setminus$ M & even & factorial & divided\_by\_three & odd & prime\\
\hline
  2 & $\times$ & $\times$ &  & & $\times$ \\
  5 &  & &  & $\times$ & $\times$ \\
  6 & $\times$ & $\times$ & $\times$ & & \\
  1 & & $\times$ & & $\times$ & $\times$ \\
  9 & & & $\times$ & $\times$ & \\
  3 & & & $\times$ & $\times$ & $\times$\\
  \hline
\end{tabular}
\end{center}
$prime, divided\_by\_three => odd$\\
Is the following implication valid?\\
The only prime that is divided\_by\_three is three itself, so it is true.
$even => factorial$\\
Is the following implication valid?\\
Not all even numbers are factorials, e.g. 8.
\begin{center}
\begin{tabular}{|c|c|c|c|c|c|}
  \hline
  G $\setminus$ M & even & factorial & divided\_by\_three & odd & prime\\
\hline
  2 & $\times$ & $\times$ &  & & $\times$ \\
  5 &  & &  & $\times$ & $\times$ \\
  6 & $\times$ & $\times$ & $\times$ & & \\
  1 & & $\times$ & & $\times$ & $\times$ \\
  9 & & & $\times$ & $\times$ & \\
  3 & & & $\times$ & $\times$ & $\times$\\
  8 & $\times$ & & & &\\
  \hline
\end{tabular}
\end{center}
$even, odd => factorial, prime, divided\_by\_three$\\
Is the following implication valid?\\
We do not have numbers which are both even and odd.
$even, divided\_by\_three => factorial$\\
Is the following implication valid?\\
We have number 12, which is even and divided\_by\_three, but it is not a factorial.
\begin{center}
\begin{tabular}{|c|c|c|c|c|c|}
  \hline
  G $\setminus$ M & even & factorial & divided\_by\_three & odd & prime\\
\hline
  2 & $\times$ & $\times$ &  & & $\times$ \\
  5 &  & &  & $\times$ & $\times$ \\
  6 & $\times$ & $\times$ & $\times$ & & \\
  1 & & $\times$ & & $\times$ & $\times$ \\
  9 & & & $\times$ & $\times$ & \\
  3 & & & $\times$ & $\times$ & $\times$\\
  8 & $\times$ & & & &\\
  12 & $\times$ & & $\times$ & &\\
  \hline
\end{tabular}
\end{center}
$even, prime => factorial$\\
Is the following implication valid?\\
The only even prime number is 2.
So, the exploration process is over.

The final context:
\begin{center}
\begin{tabular}{|c|c|c|c|c|c|}
  \hline
  G $\setminus$ M & even & factorial & divided\_by\_three & odd & prime\\
\hline
  1 & & $\times$ & & $\times$ & $\times$ \\
  2 & $\times$ & $\times$ &  & & $\times$ \\
  3 & & & $\times$ & $\times$ & $\times$\\
  5 &  & &  & $\times$ & $\times$ \\
  6 & $\times$ & $\times$ & $\times$ & & \\
  8 & $\times$ & & & &\\
  9 & & & $\times$ & $\times$ & \\
  12 & $\times$ & & $\times$ & &\\
  \hline
\end{tabular}
\end{center}
The set of implications:
\begin{itemize}
\item$factorial, odd \to prime$
\item$factorial, divided\_by\_three \to even$
\item$prime, divided\_by\_three \to odd$
\item$even, odd \to factorial, prime, divided\_by\_three$
\item$even, prime \to factorial$
\end{itemize}

\subsection{Numbers: model-based}
We can consider the same problem and use pairwise testing. One of known tools for this is PICT, pairwise testing tool by Microsoft. The initial model is quite simple.
\begin{itemize}
\item Event: 1, 0
\item Factorial: 1, 0
\item Divs3: 1, 0
\item Odd: 1, 0
\item Prime: 1, 0
\end{itemize}
The result of pairwise generation is shown in table below:
\begin{center}
\begin{tabular}{|c|c|c|c|c|}
\hline
Even & Factorial & Divs3 & Odd& Prime \\
1 & 1 & 0 & 0 & 1\\
0 & 0 & 1 & 1 & 0 \\
1 & 0& 1 & 1 & 1\\
0 & 1 & 0 & 1 & 0\\
1 & 0& 1& 0 & 0\\
0 & 1 & 1 & 0 & 1\\
0 & 0 & 0 & 0 & 0\\
\hline
\end{tabular}
\end{center}
To get the results as in the previous section, we have to modify the input model in the following way:
\begin{enumerate}
\item Parameters:
\begin{itemize}
\item Event: 1, 0
\item Factorial: 1, 0
\item Divs3: 1, 0
\item Odd: 1, 0
\item Prime: 1, 0
\item Result: 12, 9, 6, 3, 2, 1
\end{itemize}
\item Implications
\begin{itemize}
\item IF [Even] = 1 THEN [Odd] = 0 ELSE [Odd] = 1;
\item IF [Odd] = 1 AND [Factorial] = 1 THEN [Result] = 1;
\item IF [Even] = 1 AND [Prime] = 1 THEN [Result] = 2;
\item IF [Divs3] = 1 AND [Prime] = 1 THEN [Result] = 3;
\item IF [Divs3] = 1 AND [Even] = 1 THEN [Result] IN 6, 12;
\item IF [Divs3] = 1 AND [Odd] = 1 AND [Prime] = 0 THEN [Result] = 9;
\item IF [Even] = 1 AND [Factorial] = 1 AND [Divs3] = 0 THEN [Result] =
2;
\end{itemize}
\item Data-specific dependencies
\begin{itemize}
\item IF [Result] = 1 THEN [Even] = 0 AND [Factorial] = 1 AND [Divs3] = 0 AND [Odd] = 1 AND [Prime] = 1;
\item IF [Result] = 2 THEN [Even] = 1 AND [Factorial] = 1 AND [Divs3] = 0 AND [Odd] = 0 AND [Prime] = 1;
\item IF [Result] = 3 THEN [Even] = 0 AND [Factorial] = 0 AND [Divs3] = 1 AND [Odd] = 1 AND [Prime] = 1;
\item IF [Result] = 6 THEN [Even] = 1 AND [Factorial] = 1 AND [Divs3] = 1 AND [Odd] = 0 AND [Prime] = 0;
\item IF [Result] = 9 THEN [Even] = 0 AND [Factorial] = 0 AND [Divs3] = 1 AND [Odd] = 1 AND [Prime] = 0;
\item IF [Result] = 12 THEN [Even] = 1 AND [Factorial] = 0 AND [Divs3] = 1 AND [Odd] = 0 AND [Prime] = 0;
\end{itemize}
\end{enumerate}
For such model PICT outputs the following cases:
\begin{center}
\begin{tabular}{|c|c|c|c|c|c|}
\hline
Even & Factorial & Divs3 & Odd& Prime & Result \\
0 & 0 & 1 & 1 & 1 & 3\\
1 & 1& 1& 0&0&6\\
0&0&1&1&0&9\\
1&0&1&0&0&12\\
1&1&0&0&1&2\\
0&1&0&1&1&1\\
\hline

\end{tabular}

\end{center}
\subsection{Banners with geotargeting}
One of the most popular forms of advertising in the Internet is contextual advertisement. The key point is that companies pay for the click on their sites, but not for the shows, so the targeting part gains more importance. One of most-widely used features is geotargeting, i.e., defining in which regions the banner could be shown. Usually, regions are treated like a tree and if banner targets at some region, then it could be shown in it, and all it subregions.  The possible sources for defining region: query argument, search query, ip address.
Attributes with binary domains:
$M =\{IsQueryPresent(IQP), \\
IsQueryValid(IQV), IsSqPresent(ISP), IsSqValid(ISV), \\
IsIpPresent(IIP), IsIpValid(IIV), \\
UserRegion==BannerRegion(=), \\
UserRegion<BannerRegion(<), \\
UserRegion>BannerRegion(>), \\
IsBannerShown(IBS)\}$
\begin{center}
\begin{tabular}{|c|c|c|c|c|c|c|c|c|c|c|}
\hline
 & IQP & IQV & ISP& ISV & IIP &IIV & = &  < & >& IBS  \\
 \hline
1 & $\times$  & & & & & & & & & \\
2 &  & & $\times$& & & & & & & \\
3 &  & & & & $\times$& & & & & \\
4 & $\times$ &$\times$ & & & & &$\times$ & & &$\times$ \\
5 &  $\times$&$\times$ & & & & & & $\times$& & $\times$ \\
6 &  & &$\times$ &$\times$ & & &$\times$ & & &$\times$ \\
7 &  & &$\times$ &$\times$ & & & & $\times$& & $\times$\\
8 & $\times$ &$\times$ & & & & & & & $\times$& \\
9 &  & &$\times$ &$\times$ & & & & & $\times$& \\
10 &  & & & &$\times$ &$\times$ &$\times$ & & &$\times$ \\
11 &  & & & &$\times$ &$\times$ & &$\times$ & & $\times$ \\
12 &  & & & &$\times$ &$\times$ & & &$\times$ & \\
13 &$\times$  &$\times$ &$\times$ &$\times$ & & & $\times$& & &$\times$ \\
14 &$\times$  &$\times$ &$\times$ &$\times$ & & & &$\times$ & &$\times$ \\
15 &$\times$  &$\times$ &$\times$ &$\times$ & & & & &$\times$ & \\
16 &$\times$  &$\times$ & & &$\times$ &$\times$ &$\times$ & & &$\times$ \\
17 &$\times$  &$\times$ & & &$\times$ &$\times$ & &$\times$ & &$\times$ \\
18 &$\times$  &$\times$ & & &$\times$ &$\times$ & & &$\times$ & \\
19 &  & &$\times$ &$\times$ &$\times$ &$\times$ &$\times$ & & &$\times$ \\
20 &  & &$\times$ &$\times$ &$\times$ &$\times$ & &$\times$ & &$\times$ \\
21 &  & &$\times$ &$\times$ &$\times$ &$\times$ & & &$\times$ & \\
22 &$\times$  & &$\times$ & & & & & & & \\
23 &$\times$  & & & &$\times$ & & & & & \\
24 &  & &$\times$ & &$\times$ & & & & & \\
25 &$\times$  & &$\times$ & &$\times$ & & & & & \\
26 &$\times$  & &$\times$ & &$\times$ &$\times$ &$\times$ & & &$\times$ \\
27 &$\times$  &$\times$ &$\times$ & & & &$\times$ & & &$\times$ \\
28 &$\times$  &$\times$ &$\times$ & &$\times$ & &$\times$ & & &$\times$ \\
29 &$\times$  &$\times$ &$\times$ & &$\times$ &$\times$ &$\times$ & & &$\times$ \\
30 &$\times$  & &$\times$ &$\times$ & & &$\times$ & & &$\times$ \\
31 &$\times$  & &$\times$ &$\times$ &$\times$ & &$\times$ & & &$\times$ \\
32 &$\times$  & &$\times$ &$\times$ &$\times$ &$\times$ &$\times$ & & &$\times$ \\
33 &$\times$  &$\times$ &$\times$ &$\times$ &$\times$ &$\times$ &$\times$ & & &$\times$ \\
34 &$\times$  &$\times$ & & &$\times$ &$\times$ & &$\times$ & &$\times$ \\
35 &$\times$  &$\times$ & & &$\times$ &$\times$ & & &$\times$ & \\
36 & $\times$ & &$\times$ &$\times$ & & & &$\times$ & &$\times$ \\
37 & $\times$ & &$\times$ &$\times$ & & & & &$\times$ & \\
38 &$\times$  & &$\times$ & &$\times$ &$\times$ & &$\times$ & &$\times$ \\
39 &$\times$  & &$\times$ & &$\times$ &$\times$ & & &$\times$ & \\
40 &$\times$  & & & &$\times$ &$\times$ & &$\times$ & &$\times$ \\
41 &$\times$  & & & &$\times$ &$\times$ & & &$\times$ & \\
42 &  & &$\times$ & &$\times$ &$\times$ &$\times$ & & &$\times$ \\
43 &  & &$\times$ &$\times$ &$\times$ & &$\times$ & & &$\times$ \\
44 & $\times$ &$\times$ &$\times$ & &$\times$ & & &$\times$ & &$\times$ \\
45 & $\times$  &$\times$ &$\times$ & &$\times$ & & & &$\times$ & \\
46 &$\times$  & &$\times$ &$\times$ &$\times$ & & &$\times$ & &$\times$ \\
47 & $\times$ & &$\times$ &$\times$ &$\times$ & & & &$\times$ & \\
48 &$\times$  &$\times$ &$\times$ &$\times$ &$\times$ & &$\times$ & & &$\times$ \\
49 &$\times$  &$\times$ &$\times$ &$\times$ &$\times$ & & &$\times$ & &$\times$ \\
50 &$\times$  &$\times$ &$\times$ &$\times$ &$\times$ & & & &$\times$ & \\
51 &$\times$  &$\times$ &$\times$ & &$\times$ &$\times$ & &$\times$ & &$\times$ \\
52 &$\times$  &$\times$ &$\times$ & &$\times$ &$\times$ & & &$\times$ & \\
53 &$\times$  & &$\times$ &$\times$ &$\times$ &$\times$ & &$\times$ & &$\times$ \\
54 &$\times$  & &$\times$ &$\times$ &$\times$ &$\times$ & & &$\times$ & \\
55 &$\times$  &$\times$ &$\times$ &$\times$ &$\times$ &$\times$ & &$\times$ & &$\times$ \\
56 &$\times$  &$\times$ &$\times$ &$\times$ &$\times$ &$\times$ & & &$\times$ & \\
57 &  & &$\times$ & &$\times$ &$\times$ & &$\times$ & &$\times$ \\
58 &  & &$\times$ & &$\times$ &$\times$ & & &$\times$ & \\
59 &$\times$  &$\times$ & & &$\times$ & &$\times$ & & &$\times$ \\
60 &$\times$  & & & &$\times$ &$\times$ &$\times$ & & &$\times$ \\
61 &  & & & & & & & & & \\
62 &  & &$\times$ &$\times$ &$\times$ & & &$\times$ & &$\times$ \\
63 &  & &$\times$ &$\times$ &$\times$ & & & &$\times$ & \\
64 &$\times$  &$\times$ & & &$\times$ & & &$\times$ & &$\times$ \\
65 &$\times$  &$\times$ & & &$\times$ & & & &$\times$ & \\
66 &$\times$  &$\times$ &$\times$ & & & & &$\times$ & &$\times$ \\
67 &$\times$  &$\times$ &$\times$ & & & & & &$\times$ & \\
\hline

\end{tabular}

\end{center}
Implication list:
\begin{enumerate}
\item \textbf{IF} IQV \textbf{THEN} IQP
\item \textbf{IF} ISV  \textbf{THEN} ISP
\item \textbf{IF} IIV \textbf{THEN} IIP
\item \textbf{IF} = \textbf{THEN} IBS, not <, not >
\item \textbf{IF} < \textbf{THEN} IBS, not =, not >
\item \textbf{IF} IBS \textbf{THEN} not >
\item \textbf{IF} not IQP \textbf{THEN} not IQV
\item \textbf{IF} not ISP \textbf{THEN} not ISV
\item \textbf{IF} not IIP \textbf{THEN} not IIV
\item \textbf{IF} not IIV, not ISV, not IQV  \textbf{THEN} not =, not <, not >, not IBS
\item \textbf{IF} not <, not =  \textbf{THEN} not IBS
\item \textbf{IF} not >, IQV, IQP \textbf{THEN} IBS
\item \textbf{IF} not >, ISV, ISP \textbf{THEN} IBS
\item \textbf{IF} not >, IIV, IIP \textbf{THEN} IBS
\item \textbf{IF} not >, not ISV, not IQV, IBS \textbf{THEN} IIV
\item \textbf{IF} not >, not ISV, not IIV, IBS \textbf{THEN} IQV
\item \textbf{IF} not >, not IIV, not IQV, IBS \textbf{THEN} ISV
\item \textbf{IF} not >, not =, IBS \textbf{THEN} <
\item \textbf{IF} not >, not <, IBS \textbf{THEN} =
\item \textbf{IF} not IBS \textbf{THEN} not =, not <
\item \textbf{IF} not IBS, not <, not =, IQV, IQP \textbf{THEN} >
\item \textbf{IF}  not IBS, not <, not =, ISV, ISP \textbf{THEN} >
\item \textbf{IF}  not IBS, not <, not =, IIV, IIP\textbf{THEN} >
\item \textbf{IF}  not IBS, not <, not =, >, not IQV, not ISV\textbf{THEN} IIV, IIP
\item \textbf{IF}  not IBS, not <, not =, >, not IIV, not IQV\textbf{THEN} ISV, ISP
\item \textbf{IF}  not IBS, not <, not =, > not IIV, not ISV\textbf{THEN}  IQV, IQP
\item \textbf{IF}  not IBS, not >, not =, not <\textbf{THEN} not IQV, not ISV, not IIV
\end{enumerate}

\section{Plug-in setup}
The most important advantage of the proposed approach is the plug-in design. It can be easily incorporated in existing process of test-case generation. For manual test-case design one can develop test-cases in proposed system. Moreover, the step of implication extraction could be postponed up to the review of obtained cases. It is important to notice that Attribute Exploration is a good technique to extract dependencies that could be formulated as requirements for the system under test.
For the case of automatic test case generation, extraction of dependencies could eliminate the step of test debug and replace it with review of obtained implications and the model of the system under test. Also the proposed technique is applicable to verify the correctness of defining type of model, precisely wiseness of it. Since big values of $n$ in n-wise modeling impose more test cases and it results in the growing time of test run execution.

\section{Conclusion}
Formal Concept Analysis provides us with useful framework for software testing tasks. It is especially beneficial in regression testing meta report construction, feature navigation, test case analysis and derivation. It unites best practices of manual development and automatic generation of test scenarios. It provides sound and complete description of the investigated domain, based on expert knowledge. The output of the system consists of two main parts: the description of typical objects in the domain, and interdependence between parameters in terms of implications.
An important advantage of proposed technique is extensibility. If we add a new attribute, we can just copy the previous examples into new formal context, assuming that new attribute is absent for all objects and proceed with the procedure of attribute exploration. It holds even for the very start of procedure. We can start with non-empty set of objects and implications simultaneously.
The described algorithm could be used as a standalone solution for the test case design, as well as, tool to get exiting dependencies in the domain. The obtained implications could be valuable in pairwise testing to adjust the model description.

However, we should admit that the current approach is limited in terms of attribute description. For now, it is highly dependent on the boolean nature of attributes. One of the main directions of future work is to work with descriptions of general form by means of patterns structures~\cite{gk01,k13}, an extension of FCA.

\end{document}